\documentclass[conference,10pt]{IEEEtran}
\IEEEoverridecommandlockouts

\usepackage[utf8]{inputenc}
\usepackage{color,graphicx,caption,subcaption}
\usepackage[cmex10]{amsmath}
\usepackage{amsfonts}
\usepackage[english]{babel}
\usepackage{amssymb,amsmath,mathtools}
\usepackage{enumerate}
\usepackage{algorithm}
\usepackage{algorithmic}
\usepackage{bbm}
\usepackage{amsmath}
\usepackage{mathrsfs}
\usepackage{mathtools}
\usepackage{breqn}
\usepackage[english]{babel}
\usepackage{amsthm}
\usepackage{amsmath}
\usepackage{enumitem}
\usepackage{booktabs}
\usepackage{steinmetz}
\usepackage{cite}
\usepackage{color}
\usepackage{booktabs}

\newtheorem{Lemma}{Lemma}

\newcommand{\bfh}[1]{\mathbf{#1}^{H}}
\newcommand{\bff}[1]{\mathbf{#1}}

\newcommand{\sqproof}{\hspace*{0em plus 1fill}\makebox{\hfill\ensuremath{\square}}}
\DeclareMathOperator*{\argmin}{argmin}
\DeclareMathOperator*{\argmax}{argmax}

\begin{document}
\bstctlcite{IEEEexample:BSTcontrol}
\title{Adaptive Hybrid Beamforming with Massive Phased Arrays in Macro-Cellular Networks}

\author{Shahram Shahsavari$^\dagger$, S. Amir Hosseini$^\diamond$, Chris Ng$^\diamond$, and Elza Erkip$^\dagger$
        \\ $^\dagger$ECE Department of New York University, NYU Tandon School of Engineering, New York, USA
        \\$^\diamond$Blue Danube Systems, Warren, New Jersey, USA
        \\ $^\dagger$\{shahram.shahsavari, elza\}@nyu.edu, 
        $^\diamond$\{amir.hosseini, chris.ng\}@bluedanube.com 
        
}


\maketitle

\begin{abstract}
Hybrid beamforming via large antenna arrays has shown a great potential for increasing data rate in cellular networks by delivering multiple data streams simultaneously. In this paper, several beamforming design algorithms are proposed based on the long-term channel information for macro-cellular environments where the base station is equipped with a massive phased array under per-antenna power constraint. Using an adaptive scheme, beamforming vectors are updated whenever the long-term channel information changes. First, the problem is studied when the base station has a single RF chain (single-beam scenario). Semi-definite relaxation (SDR) with randomization is used to solve the problem. As a second approach, a low-complexity heuristic beam composition algorithm is proposed which performs very close to the upper-bound obtained by SDR. Next, the problem is studied for a generic number of RF chains (multi-beam scenario) where the Gradient Projection method is used to obtain local solutions. Numerical results reveal that using massive antenna arrays with optimized beamforming vectors can lead to 5X network throughput improvement over systems with conventional antennas.
\end{abstract}


\IEEEpeerreviewmaketitle

\section{Introduction}



In light of the rapid development of fifth generation cellular networks (5G), Massive MIMO has proven to improve the network performance significantly \cite{marzetta2016fundamentals}. These systems comprise of an array of many antenna elements. 
The user data is precoded in the digital domain first and then, each of the digital streams is converted to a radio frequency signal through a circuit referred to as RF chain. Each signal is then transmitted by the antenna element connected to that RF chain. This process is best suited to a rich scattering propagation environment that provides a large number of degrees of freedom. In macro-cellular environment, however, these conditions often do not hold. A more efficient alternative is the use of hybrid massive MIMO systems in such scenarios \cite{molisch2017hybrid}.  

In hybrid Massive MIMO systems, there are fewer RF chains than antenna elements. This helps the overall system to be much less power hungry and more cost effective, since each RF chain consists of power consuming and expensive elements such as A/D and D/A converters which do not follow Moore's law. However, these systems rely on accurate channel estimation and typically are applied to TDD networks to alleviate the estimation overhead \cite{molisch2017hybrid}. On the other hand, common deployment of LTE in North America is FDD based. In this paper, we focus on a class of hybrid massive MIMO systems where all antenna elements maintain RF coherency \cite{HDAAS}. This means that all antenna elements are closely spaced and have matching phase and magnitude characteristics at the operating frequency \cite{lo2013antenna}. Using this technique, applicable also in FDD with existing LTE protocols, the antenna system can be used as a phased array and macro-cellular transmission is achieved through hybrid beamforming (BF) \cite{orfanidis2002electromagnetic}.

In hybrid BF, each RF chain carries a stream of data and is connected to each antenna element through a separate pair of variable gain amplifier and phase shifter. By setting the values of the amplifier and phase shifts (equivalently designing BF vectors), multiple beams are generated, each carrying one data stream over the air. Generating beams using phased arrays generally requires channel information of all users. By keeping the beam pattern constant over an extended period of time, e.g., one hour, small scale channel variations can be averaged out. Hence, the BF direction corresponds to a dominant multipath component \cite{li2010mimo} which mainly depends on the user location in macro-cellular environment due to the primarily LOS channels. Whenever user location information is updated, the system can adaptively switch to a different beam pattern to constantly provide enhanced service to the users. We refer to this technique as \emph{long-term adaptive BF}.

The radiated power from an antenna array is constrained and power constraints are chosen to limit the non-linear effects of the amplifiers \cite{joung2015survey}. Generally, two types of power constraints are considered in research problems: \textit{i}) sum power constraint (SPC) in which an upper-bound is considered for the total power consumption of the array, and \textit{ii}) per-antenna power constraint (PAPC) in which an upper-bound is considered for the power consumption of each antenna in the array \cite{yu2007transmitter,ng2010linear}. Although it is more convenient to consider SPC for research problems \cite{sidiropoulosULA,gershman2010convex}, it is not applicable to practical implementations, where each antenna element is equipped with a separate power amplifier. 

Generating adaptive beams that maximize the overall network throughput plays a significant role in exploiting the benefits of hybrid BF in a cellular system. Any method that is proposed should have a manageable complexity and operate within the power constraints of the array. The goal of this paper is to propose methods for long-term adaptive BF under PAPC to maximize the average network rate using hybrid phased arrays with arbitrary number of beams. 
First, we focus the optimization on an individual cell where the interference from other cells is treated as noise. We use well-known theoretical and numerical techniques for finding the optimal beam pattern as well as a theoretical upper bound for the solution. Then, we propose a low complexity heuristic algorithm that performs close to the obtained upper bound.    

\textit{Notation}: We use uppercase bold letters to denote matrices and lowercase bold letters to denote column vectors. $X_{mm}$ and $w_m$ are the $(m,m)^{th}$ and $m^{th}$ element of matrix $\bff{X}$ and vector $\bff{w}$, respectively. $(.)^T$, $(.)^H$, $\operatorname{Tr}\{.\}$, and $||.||_{\text{F}}$ are transpose, hermitian, trace and Frobenius norm operations, respectively. $[N]$ denotes the set of integers from 1 to $N$.



\section{problem statement}
\label{sec:problem-statement}
\begin{figure}[t]
 \centering \includegraphics[width=0.4\linewidth]{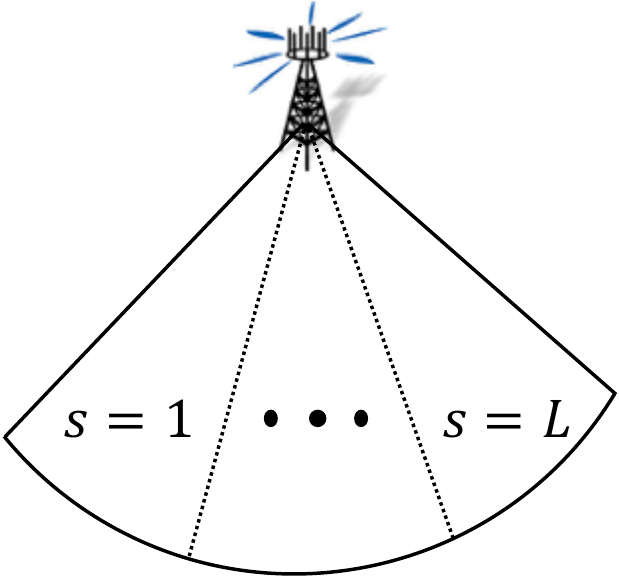}
 \caption{Single cell scenario with $L$ sections}
 \label{fig:sector}
\end{figure}

We consider the downlink of a single-cell scenario consisting of a BS with $M$ antennas and $L\ll M$ radio frequency (RF) transceiver chains \cite{molisch2017hybrid}. Since each RF chain can carry a single data stream, the BS can serve $L$ User Equipments (UEs) simultaneously. As a result, the cell site is partitioned into $L$ sections (Fig. \ref{fig:sector}) and one UE per section is activated at each time slot as will be explained later. We assume that user equipments (UEs) are clustered in hotspots within the cell. Let $H_{si}$, $(s,i)\in[L]\times[K_s]$ denote hotspot $i$ from section $s$ which consists of a group of $N_{si}$ nearby UEs. We let $K_s$ be the number of hotspots in section $s$ and $K=\sum_s K_s$ denotes the total number of hotspots in the cell. The fraction of UEs located at $H_{si}$ among the UEs in section $s$ is defined by $\alpha_{si}= N_{si}/N_s$, where $N_s=\sum_i N_{si}$ is the total number of UEs in section $s$. Let $U^n_{si}$, $n\in [N_{si}]$, denote the $n^{th}$ UE of hotspot $H_{si}$. 

We consider a macro-cellular environment in which the channels are primarily LOS with the possibility of having local scatterers around the UEs. We assume that only the long-term channel state information of UEs is available at the BS and can be used to perform long-term BF. Furthermore, we assume that the long-term channel vectors between the BS and the users belonging to a hotspot are the same due to their proximity. Let $\bff{g}_{si}=\sqrt{\beta_{si}} \bff{h}_{si}$, denote the long-term channel vector between the BS and the UEs located at $H_{si}$, where $\beta_{si}\in \mathbb{R^+}$ and $\bff{h}_{si}\in \mathbb{C}^M$ denote the pathloss and spatial signature between BS and $H_{si}$, respectively. We consider the Vandermonde model where $\bff{h}_{si}=[e^{j\theta_{si}},e^{j2\theta_{si}},\ldots, e^{jM\theta_{si}}]^T$. A use case of this channel model is when the users are located in the far-field of a uniform linear array with $M$ antennas in a primarily line-of-sight environment \cite{sidiropoulosULA}. In such cases we have $\theta_{si}= 2 \pi d \sin(\psi_{si})/ \lambda$, where $d$ denotes the spacing between successive elements, $\lambda$ is the wavelength, and $\psi_{si}$ is direction of $H_{si}$ relative to the BS. In order to model other types of antenna arrays such as rectangular and circular arrays, $\bff{h}_{si}$ can be changed accordingly. We note that this model relates the long-term channel information to the location of the hotspots. In \cite{vandermond-verification}, the validity of this model is demonstrated using a variety of test-bed experiments.

Each RF chain is connected to each antenna element through a separate pair of variable gain amplifier and phase shifter. We model the corresponding gain and phase shift by a complex coefficient. As a result, there are $M$ complex coefficients corresponding to each RF chain creating a BF vector. The radiation pattern (or equivalently beam pattern) of the antenna array corresponding to each RF chain can be modified by controlling the corresponding BF vector \cite{orfanidis2002electromagnetic}. We assume that the BS uses BF vector $\bff{w}_s\in \mathbb{C}^M,~s\in[L]$ to generate a beam pattern (or a beam in short) for serving UEs located in section $s$. To reduce the complexity, these BF vectors are designed based on the long-term channel information and are adaptively modified when the long-term channel information changes, i.e., when there is substantial change in the geographical distribution of the hotspots. Moreover, we assume that the UEs are scheduled within each section based on a round robin scheduler. Let $q^n_{si} \in \mathbb{C}$ denote the signal to be sent to $U^n_{si}$, where $\mathbb{E}(s^n_{si})=0$ and $\mathbb{E}(|q^n_{si}|^2)=1$. Also, let $U^{n^*}_{si^*}$ be the scheduled UE in section $s$ at a generic time slot. Hence, the BS transmit vector is $\bff{x}=\sqrt{P}\sum_{s\in [L]} q^{n^*}_{si^*}\bff{w}_s$, where $P$ denotes the average transmit power of the BS. Subsequently, $U^{n^*}_{si^*}$ receives signal 
\begin{align*}
y^{n^*}_{si^*}=\sqrt{P\beta_{si^*}}q^{n^*}_{si^*}\bfh{h}_{si^*}\bff{w}_s+ \sum_{s'\not=s}\sqrt{P\beta_{s'i^*}}q^{n^*}_{s'i^*}\bfh{h}_{si^*}\bff{w}_{s'}+v,
\end{align*}
where $v\sim \mathcal{CN}(0,\sigma^2)$ is the noise. The first term corresponds to the desired signal received from beam $s$ and the second term is the interference received from other $L-1$ beams. Therefore, whenever $U^{n}_{si}$ is scheduled, the corresponding SINR is 
\begin{align}
\text{SINR}_{si}(\bff{W})=\frac{\bfh{w}_{s} \bff{Q}_{si} \bff{w}_{s}}{1+\sum_{s'\not=s}\bfh{w}_{s'} \bff{Q}_{si}\bff{w}_{s'}},
\end{align}
where $\bff{Q}_{si}=\gamma_{si}\bff{h}_{si}\bfh{h}_{si}$ with $\gamma_{si}=P\beta_{si}/\sigma^2$. The BF matrix is defined as $\bff{W}\triangleq[\bff{w}_1,\bff{w}_2,\ldots,\bff{w}_L]\in \mathbb{C}^{M\times L}$ and has columns corresponding to BF vectors of different sections. We note that PAPC corresponds to $\sum_{s\in [L]}|W_{ms}|^2 \leq 1/M, \forall m$. Hence we define the feasible set as $\mathcal{A}=\{\bff{W} \in \mathbb{C}^{M\times L}|~\forall m: \sum_{s\in [L]}|W_{ms}|^2 \leq 1/M \}$. 
The goal is to find BF matrix $\bff{W}$ which maximizes a network utility function, denoted by $R(\bff{W})$ over the feasible set $\mathcal{A}$. In this paper, we consider average network rate as the network utility, i.e., 
\begin{align}
R(\bff{W})=\sum_{s\in[L]}\sum_{i\in[K_s]} \alpha_{si} \log (1+\text{SINR}_{si}).
\end{align}
Hence, The problem can be formulated as follows. 
\begin{align}
&\Pi_L:~\bff{W}^{opt}=\argmax_{\bff{W}\in \mathcal{A}}~R(\bff{W}) \notag
\end{align}
We note that the sub-index $L$ in $\Pi_L$ corresponds to the number of the beams (equivalently number of RF chains). Although there is no minimum utility constraint defined for individual UEs in problem $\Pi_L$, sum-log maximization induces a type of proportional fairness. It can be shown that problem $\Pi_L$ is not in a convex form \cite[chapters 3, 4]{boyd2004convex}.
Therefore, finding the globally optimal solution of this problem is difficult. In section \ref{sec:single-beam}, we study the single-beam ($L=1$) problem $\Pi_1$ to find local solutions and an upper-bound to evaluate the performance. In section \ref{sec:multi-beam}, we provide an iterative algorithm to find a sub-optimal solution of problem $\Pi_L$ for arbitrary $L$.

In Sections \ref{sec:single-beam} and \ref{sec:multi-beam}, we will need to find the \textit{projection} of a general beamforming matrix $\bff{W} \in \mathbb{C}^{M\times L}$ on set $\mathcal{A}$ which is defined as
\begin{align}
\mathbb{P}_{\mathcal{A}}(\bff{W})=\argmin_{\bff{X}\in \mathcal{A}}~ ||\bff{X}-\bff{W}||^2_{\text{F}}. \label{eq:optimal-projection}
\end{align}
We note that $\mathcal{A}$ is a closed convex set which leads to a unique $\mathbb{P}_{\mathcal{A}}(\bff{W})$ for every $\bff{W} \in \mathbb{C}^{M\times L}$, introduced by Lemma \ref{lem:optimal-projection}. The proof is provided in Apeendix \ref{sec:app-lemma1}.
\begin{Lemma}
\label{lem:optimal-projection}
We have $\bff{\hat{W}}=\mathbb{P}_{\mathcal{A}}(\bff{W})$ if and only if for every $m \in \{1,2,\ldots,M\}$ 
\begin{align}
\begin{cases}
\hat{W}_{ms}=W_{ms}& \mbox{if } \sum_s|W_{ms}|^2 \leq 1/M,\\ 
\hat{W}_{ms}=W_{ms}/(\sqrt{M\sum_s|W_{ms}|^2}) & \mbox{if } \sum_s|W_{ms}|^2>1/M.
\end{cases}\notag
\end{align}
\end{Lemma}


\section{Single-Beam Scenario}
\label{sec:single-beam}
In this section, we study problem $\Pi_1$ where every UE is served by a single-beam generated by BF vector $\bff{w}$, i.e., there is only one section in the cell ($s=1$) and PAPC corresponds to $|w_m|^2\leq 1/M,\forall m$. In this scenario, there is no interference since one UE is scheduled per time slot. Hence we have $R(\bff{w})=\sum_{i\in[K]} \alpha_{i} \log (1+\bfh{w}\bff{Q}_{i} \bff{w})$. Please note that we drop index $s$ in the single-beam scenario because $s=1$. Next, we derive an upper-bound for the optimal value of $\Pi_1$ and provide two different methods to obtain local solutions of this problem. We will use the upper-bound as a benchmark in the simulations to evaluate the effectiveness of the local solutions. 
\subsection{Semi-definite relaxation with randomization}
\label{subsec:SDR}
Since $\bfh{w}\bff{Q}_{i} \bff{w}$ is a complex scalar, we have $\bfh{w}\bff{Q}_{i} \bff{w}=(\bfh{w}\bff{Q}_{i} \bff{w})^T=\gamma_i\bfh{h}_i\bff{X}\bff{h}_i$, where $\bff{X}=\bff{w}\bfh{w}\in \mathbb{C}^{M\times M}$ is a rank-one  positive semi-definite matrix. Using this transformation, semi-definite relaxation of problem $\Pi_{1}$ is as follows.
\begin{align}
&\boldsymbol{\Pi}_{1r}:~ \bff{X}_r^{opt}=\argmax_{\bff{X}\in \mathbb{C}^{M\times M}}~ \sum_{i=1}^K \alpha_i \log(1+\gamma_i \bfh{h}_i\bff{X}\bff{h}_i) \notag \\
&\text{subject to:}~~  \bff{X}\geq 0,~\forall m: X_{mm} \leq 1/M \notag 
\end{align}
We remark that $\Pi_{1}$ is equivalent to $\Pi_{1r}$ plus a non-convex constraint $Rank(\bff{X})=1$. Removing the rank-one constraint enlarges the feasible set and makes it possible to find solutions with higher objective value. Hence, the optimal objective value of $\Pi_{1r}$ is an upper-bound for the optimal objective value of $\Pi_1$. This is a well-known technique called `Semi-definite Relaxation (SDR)' \cite{luo2010semidefinite}. Note that $\Pi_{1r}$ can be solved using convex programming techniques \cite{boyd2004convex}. After solving the convex problem $\Pi_{1r}$ there are two possibilities: 
\begin{enumerate}[leftmargin=*]
\item {$\bf Rank(\bff{X}_r^{opt})=1$}: 
in this case the upper-bound is tight and we have $\bff{X}_r^{opt}=\bff{w}^{opt}{\bff{w}^{opt}}^H$, where $\bff{w}^{opt}$ is the solution of $\Pi_1$.

\item {$\bf Rank(\bff{X}_r^{opt})>1$}:
in this case, the upper-bound is not tight and finding the global solution of $\Pi_1$ is difficult. However, there are a number of methods developed to generate a reasonable BF vector $\bff{w}$ for problem $\Pi_1$ by processing $\bff{X}_r^{opt}$ \cite{luo2010semidefinite}. For example, using eigenvalue decomposition, we have $\bff{X}_r^{opt}=\bff{V} \bff{\Lambda} \bff{V}^H$. Let $\bff{v}_1,\bff{v}_2,\ldots,\bff{v}_M$ be the eigenvectors in descending order of eigenvalues. One simple approach is to use the eigenvector corresponding to the maximum eigenvalue and form BF vector as $\bff{w}_{mev}=\frac{\bff{v}_1}{\sqrt{M}||\bff{v}_1||_2}$. It should be noted that normalization is necessary for feasibility. Although this simple method is optimal when $Rank(\bff{X}_r^{opt})=1$, it is not the best strategy when $Rank(\bff{X}_r^{opt})>1$. Using different `\textit{randomization}' techniques can lead to better solutions \cite{luo2010semidefinite}. 
Let us define $\bff{w}_{sdr}=\frac{\bff{b}}{\sqrt{M}||\bff{b}||_2}$, where $\bff{b}=\bff{V}\bff{\Lambda}^{1/2}\bff{e}$ with random vector $\bff{e}\in \mathbb{C}^M$. The elements of $\bff{e}$ are \textit{i.i.d.} random variables uniformly distributed on the unit circle in the complex plane. 
Alternative distributions such as Gaussian distribution can also be adopted for $\bff{e}$\cite{sidiropoulos2006transmit}. The randomization method is to generate a number of BF vectors $\{\bff{w}_{sdr}\}$ and pick the one resulting in the highest objective value of $\Pi_1$. Note that using $\bff{e}=[1,0,0,\ldots,0]^T$ would lead to $\bff{w}_{sdr}=\bff{w}_{mev}$. The number of random instances denoted by $N_{trial}$ depends on the number of the hotspots, which is discussed more in the numerical examples. 
\end{enumerate}

\subsection{Single-beam sub-beam composition}
\label{subsec:sbc}

In this section, we introduce a heuristic algorithm to find a BF vector $\bff{w}$ for $\Pi_1$ with a relatively good performance compared to the upper-bound obtained by SDR. Suppose there is only one hotspot in the network, say $H_i$. Using Cauchy-Schwarz inequality, we can show that the solution of $\Pi_1$ is $\bff{w}\triangleq \bff{h}_i/\sqrt{M}$. In this technique, which is referred to as \textit{conjugate beamforming}, the BS creates a narrow beam towards the location of $H_i$ \cite[chapter 19]{orfanidis2002electromagnetic}. We can generalize this method to generate a beam pattern serving all the hotspots, by summing up the individually optimal BF vectors, and normalizing the result to satisfy PAPC. Hence, the resulting BF vector is  $\bff{w}_{sbc}\triangleq \mathbb{P}_{\mathcal{A}}(\sum_{i=1}^K \bff{w}_i)$, where $\mathbb{P}_{\mathcal{A}}(.)$ is given by Lemma \ref{lem:optimal-projection}. We call this method \textit{single-beam sub-beam composition (SB-SBC)} due to the fact that we form a beam pattern by adding up multiple sub-beams. 

Adding up individually optimal BF vectors and projecting the result on the feasible set $\mathcal{A}$ will perturb each of them. Therefore, $\bff{w}_{sbc}$ would not exactly point towards all the hotspots. To compensate for this disturbance, we use another approach called single-beam phase optimized sub-beam composition (SB-POSBC). In SB-POSBC, we add a separate phase shift for each BF vector $\bff{w}_i$ in the summation, i.e., we define $\bff{w}_{posbc}\triangleq \mathbb{P}_{\mathcal{A}}(\sum_{i=1}^K e^{j\phi_i}\bff{w}_i)$. By choosing a set of appropriate phase shifts, $\bff{w}_{posbc}$ leads to a beam pattern which points to all the hotspots, hence, it leads to a better network utility. Since it is not easy to find optimal phase shifts analytically, one approach is to try a number of randomly chosen sets of phase shifts and pick the one which leads to the highest objective value in $\Pi_1$. One can think of this random trials as the counterpart of randomization technique described in section \ref{subsec:SDR}. Note that if $\forall i: \phi_i=0$ then $\bff{w}_{posbc}=\bff{w}_{sbc}$, hence, if the case of zero phase shifts is included in the set of random phase shifts, we can ensure that SB-POSBC will perform at least as good as SB-SBC. One important parameter in SB-POSBC is the number of random trials of phase shift sets denoted by $N_{trial}$ which will be studied in section \ref{subsec:trials}. 


\begin{figure}[t]
 \centering \includegraphics[width=0.9\linewidth]{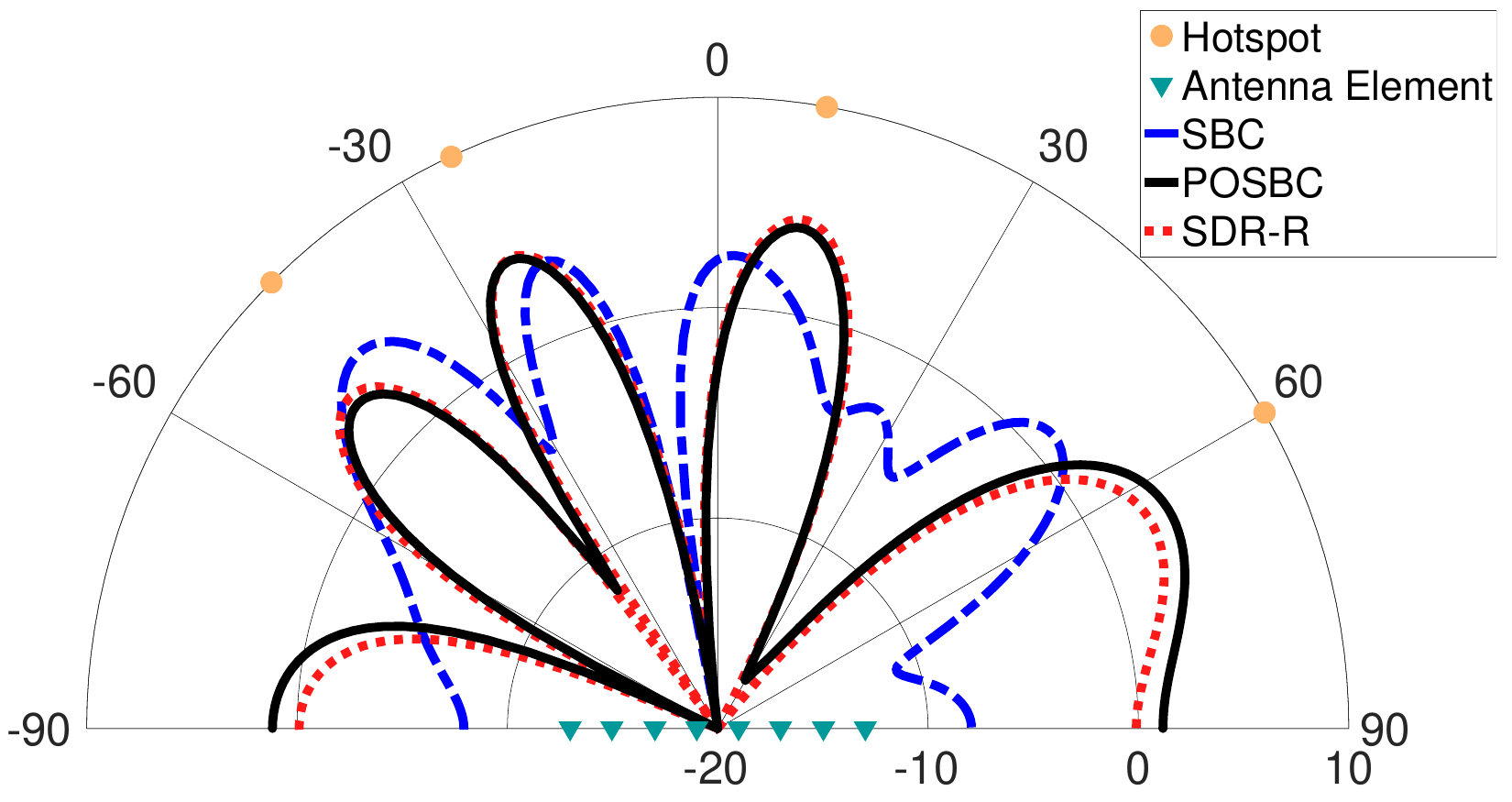}
 \caption{Comparison between the beam patterns (in dB) generated by SB-SBC and SB-POSBC for a uniform linear antenna array with eight antennas (8-ULA) and four hotspots.}
 \label{fig:sample_sbc}
\end{figure}
Figure \ref{fig:sample_sbc} depicts a network with four hotspots. This figure also depicts the beam patterns corresponding to BF vectors $\bff{w}_{sbc}$, $\bff{w}_{posbc}$, and $\bff{w}_{sdr}$ with $N_{trial}=1000$. We can observe how phase shifts in SB-POSBC compensate for the perturbation caused by SB-SBC. Furthermore, we can also see that SB-POSBC creates a similar beam to SDR with randomization while its complexity is much lower. 
\section{Multi-Beam Scenario}
\label{sec:multi-beam}
In this section, we study problem $\Pi_L$ for generic $L$. First we present a heuristic similar to SB-SBC and SB-POSBC, described in Section \ref{subsec:sbc}, and then we introduce an iterative algorithm to find a local solution of problem $\Pi_L$.
\subsection{Multi-beam sub-beam composition}
\label{subsec:mb-sbc}
Similar to what is described in Section \ref{subsec:sbc}, one can obtain $L$ BF vectors each of which generates a beam to cover a section. To this end, we can consider each section and its associated hotspots and use SB-SBC (or SB-POSBC) to find a BF vector for that section. Furthermore, we assume that the power is equally divided among the BF vectors. Hence, after applying SB-SBC (or SB-POSBC) to find a BF vector for each section separately, we divide all the vectors by $1/\sqrt{L}$. We call this method MB-SBC (or MB-POSBC). We note that this method does not consider inter-beam interference, because each BF vector is obtained independently from the others.
\subsection{Gradient projection}
Numerical optimization methods can be used to find a local solution of $\Pi_L$ for arbitrary $L$. These methods are more valuable when it is difficult to find a closed-form solution, such as non-convex non-linear optimization. Although there is no guarantee that these methods find the global optimum, they converge to a local optimum of if some conditions hold; we refer the reader to \cite{bertsekas1999nonlinear} for details. To find a local solution for problem $\Pi_L$, we use an iterative numerical method called `Gradient Projection (GP)'. 
Although there are different types of GP, we use one that includes two steps at each iteration: \textit{i}) taking a step in the gradient direction of the objective function with a step-size satisfying a condition called Armijo Rule (AR), and \textit{ii}) projecting the new point on the feasible set. 
Let $\bff{W}^{[k]}$ be the BF matrix at iteration $k$. We define
\begin{align}
&\bff{W}^{[k+1]}=\mathbb{P}_{\mathcal{A}}\big( \bff{W}^{[k]}+r^{[k]}\bff{G}^{[k]} \big), \label{eq:gp-update}
\end{align}
where $\bff{G}^{[k]}=[\bff{g}_1^{[k]},\bff{g}_2^{[k]},\ldots,\bff{g}_L^{[k]}]$  with $\bff{g}^{[k]}_s\triangleq\nabla_{\bff{w}_s} R\big(\bff{W}^{[k]}\big)$, $r^{[k]}>0$. $r^{[k]}$ denotes the step-size at iteration $k$ and $\mathbb{P}_{\mathcal{A}}(\bff{W})$ is the projection of BF matrix $\bff{W}$ on the feasible set $\mathcal{A}$ which is given by Lemma \ref{lem:optimal-projection}.
We observe that the projection rule is relatively simple and does not impose high implementation complexity to the problem. 

The step-size calculation rule directly affects the convergence of GP. Applying AR to problem $\Pi_L$, we have $r^{[k]}=\tilde{r}\beta^{l^{[k]}}$ where $\tilde{r}>0$ is a fixed scalar and $l^{[k]}$ is the smallest non-negative integer satisfying $R(\bff{W}^{[k+1]})-R(\bff{W}^{[k]})\geq \sigma \operatorname{Re}\big[\operatorname{Tr}\{(\bff{W}^{[k+1]}-\bff{W}^{[k]})^H\bff{G}^{[k]}\}\big]$ and $\bff{W}^{[k+1]}$ is given by \eqref{eq:gp-update}. In order to find $r^{[k]}$ at iteration $k$, we start from $l^{[k]}=0$ and increase $l^{[k]}$ one unit at a time until the above condition is satisfied. $0<\sigma<1$ and $0<\beta<1$ are AR parameters. In practice, $\sigma$ is usually chosen close to zero, e.g., $\sigma \in [10^{-5},10^{-1}]$. Also, $\beta$ is usually chosen between $0.1$ and $0.5$ \cite{bertsekas1999nonlinear}.

\begin{Lemma}
\label{lem:gp-convergence}
Let $\{\bff{W}^{[k]}\}$ be a sequence of BF matrices generated by gradient projection in \eqref{eq:gp-update} with step-size $r^{[k]}$ chosen by the Armijo rule, described above. Then, every limit point of $\{\bff{W}^{[k]}\}$ is stationary.
\end{Lemma}
We refer the reader to \cite[chapter 2]{bertsekas1999nonlinear} for the proof. To implement GP, we need an initial point $\bff{W}^{[1]}$ and a termination condition. We use MB-SBC described in Section \ref{subsec:mb-sbc} to generate an initial point for the numerical examples. For the termination condition, we define the error as $err^{[k+1]}\triangleq ||\bff{W}^{[k+1]}-\bff{W}^{[k]}||_{\text{F}}$ and we stop after iteration $k$ if $err^{[k+1]}\leq\epsilon$, where $\epsilon$ is a predefined error threshold. Although the numerical examples will show that GP converges fast with AR, we specify a threshold for the number of iterations denoted by $N_{iter}$ to avoid slow convergence. 

\begin{figure}[t]
 \centering \includegraphics[width=0.9\linewidth]{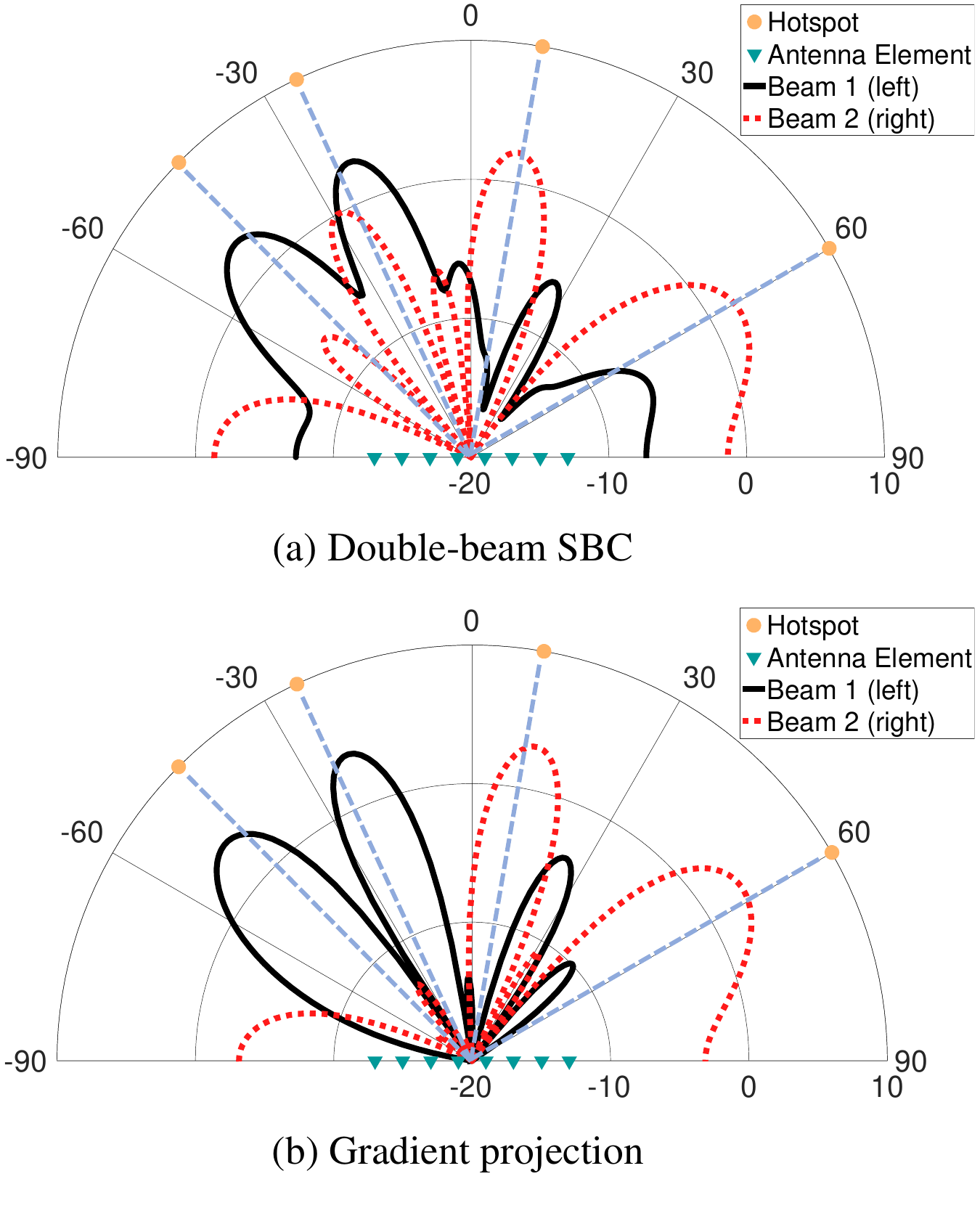}
 \caption{Beam patterns generated by double-beam SBC and gradient projection in a sample network with 2 beams, four hotspots, and a uniform linear array with eight antennas.}
 \label{fig:sample_double}
\end{figure}

Fig. \ref{fig:sample_double} illustrates a network with two beams where sections 1 and 2 are the left and right half planes, respectively, and each beam serves 2 hotspots. Fig. \ref{fig:sample_double}a shows the beams generated by double-beam SBC described in Section \ref{subsec:mb-sbc}. We observe that the BS suffers from inter-beam interference in this case. GP takes the solution of double-beam SBC as initial point and iteratively updates the BF coefficients of each beam (Fig. \ref{fig:sample_double}b), which greatly reduces the inter-beam interference.

\section{Numerical examples}\label{sec:numerical-results}
In this section, we provide numerical examples to evaluate and compare the performance of the proposed methods. We simulate the downlink of a three dimensional network with a BS consisting of a $4 \times 12$ uniform rectangular array serving a $120^{\circ}$ sector of a cell. Table \ref{tab:sim-param} lists the network parameters. Hotspots are distributed uniformly at random in a ring around the BS with inner and outer radii of 300 m and 577 m, respectively. We use CVX package \cite{cvx} to solve the convex problem $\Pi_{1r}$. We also use $\epsilon=10^{-4}$ and $N_{iter}=10^4$ for GP. 

\begin{table}[h]
\centering
\scalebox{1}{
\begin{tabular}{ll} \toprule
$\bf Parameter$ 				 & $\bf Value$             \\ \midrule
Scenario                         & single-beam, double-beam\\
Cell radius        				 & $577$ m                \\
Bandwidth      					 & $20$ MHz                      \\
Noise spectral density     	     & $-174$ dBm/Hz                  \\ 
BS transmit power ($P$)          & $20$ dBm                 \\
Number of hotspots ($K$)         & $4,8,16$                      \\ 
Pathloss in dB ($\beta^{-1}$)    & $128.1+37.6\log_{10}(d~\text{in km})$  \\ \bottomrule
\end{tabular}
}
\caption{Simulation parameters}
\label{tab:sim-param}
\end{table}

\label{sec:numerical-result}
\subsection{Effect of number of trials on the performance}
\label{subsec:trials}
In this section, we consider the single-beam scenario described in Section \ref{sec:single-beam}. We focus on SDR with randomization (SDR-R) and SB-POSBC described in Sections \ref{subsec:SDR} and \ref{subsec:sbc}, respectively. In both of these algorithms there are $N_{trial}$ random trials. To evaluate the performance of these algorithms, we consider 100 random network realizations. We run both algorithms with $N_{trial}=10^0,10^1,10^2,10^3,10^4$ and find the BF vectors $\bff{w}_{sdr}$ and $\bff{w}_{posbc}$ and the corresponding network utilities in bps/Hz. We also obtain the Upper-Bound (UB) by solving the relaxed problem $\Pi_{1r}$. Table \ref{tab:trial} lists the average performance of these algorithms given two values for number of hotspots, $K$. We observe that the larger $N_{trial}$ becomes, the closer the performance gets to the UB, which in turn slows down the pace of improvement. For larger $K$, however, the performance keeps increasing with the number of trials, which suggests that the number of trials should be proportional to the number of hotspots. While both algorithms provide performance close to the UB with large enough $N_{trial}$, SDR-R outperforms SB-POSBC in some cases. This improved performance comes at the cost of higher computational complexity.
\begin{table}[h]
\centering
\scalebox{1}{
\begin{tabular}{cllllll} \toprule
$\bf K$  & $\bf Method$   &               &             &  $\bf N_{trial}$&               &             \\ \midrule
         &                & $\bf 10^0$    & $\bf 10^1$  &  $\bf 10^2$     &    $\bf 10^3$ &   $\bf 10^4$\\ \cline{3-7}\\
         & $\bf SB-POSBC$    & $5.039$       & $5.505$     &  $5.602$        &    $5.627$    &   $5.637$   \\ 
$\bf 4$  & $\bf SDR-R$    & $5.252$       & $5.584$     &  $5.609$        &    $5.611$    &   $5.611$   \\
         & $\bf UB$       & $5.783$       & $5.783$     &  $5.783$        &    $5.783$    &   $5.783$   \\ \midrule
         & $\bf SB-POSBC$    & $3.617$       & $4.220$     &  $4.403$        &    $4.498$    &   $4.556$   \\ 
$\bf 16$ & $\bf SDR-R$    & $4.123$       & $4.521$     &  $4.572$        &    $4.585$    &   $4.587$   \\
         & $\bf UB$       & $4.715$       & $4.715$     &  $4.715$        &    $4.715$    &   $4.715$   \\ \bottomrule
\end{tabular}
}
\caption{Average network utility in (bps/Hz)}
\label{tab:trial}
\end{table}

\subsection{Computational complexity and performance}
\label{subsec:time-complexity}
In this part, we compare the performance and numerical complexity of different algorithms devised for the single-beam scenario ($L=1$). Table \ref{tab:run-time} lists the average throughput (in bps/Hz) and average time spent on a typical desktop computer (3.1 GHz Core i5 CPU, 16 GB RAM) to find the BF vector using GP, SB-POSBC, and SDR-R algorithms. The reported values are average values over 100 random network realizations. It is assumed that $N_{trial}=10^3$ for SDR-R and SB-POSBC. While GP and SB-POSBC have sub-second runtime, SDR-R has a much higher complexity. This is because a complex convex optimization problem has to be solved in the first step of SDR-R, whereas GP and SB-POSBC only require simple mathematical operations. We also observe that the performance of these algorithms are very close. Overall, we can conclude that GP and SB-POSBC are superior to SDR-R since they achieve similar performance with much lower computational complexity. The results also reveal that network utility decreases for each of the algorithms when the number of hotspots increases. This is the cost of having a single beam pattern. In fact, given a fixed antenna array aperture, it is more difficult to provide good BF gain for larger number of hotspots with a single beam.

\begin{table}[h]
\centering
\scalebox{1}{
\begin{tabular}{clcc} \toprule
$\bf K$  & $\bf Method$   & $\bf Run~Time$ & $\bf Network~Utility$  \\ \midrule
         & $\bf GP$       & $0.024$        & $5.541$				\\
$\bf 4$  & $\bf SB-POSBC$    & $0.081$   	   & $5.627$                \\
     	 & $\bf SDR-R$    & $9.268$        & $5.611$                \\ \midrule
         & $\bf GP$       & $0.040$        & $4.958$                \\
$\bf 16$ & $\bf SB-POSBC$    & $0.293$        & $4.498$                \\ 
     	 & $\bf SDR-R$    & $21.026$       & $4.585$                \\ \bottomrule

\end{tabular}
}
\caption{Average run time (in seconds) and utility in (bps/Hz)}
\label{tab:run-time}
\end{table}

\begin{figure}[t]
\centering 
\includegraphics[width=1\linewidth]{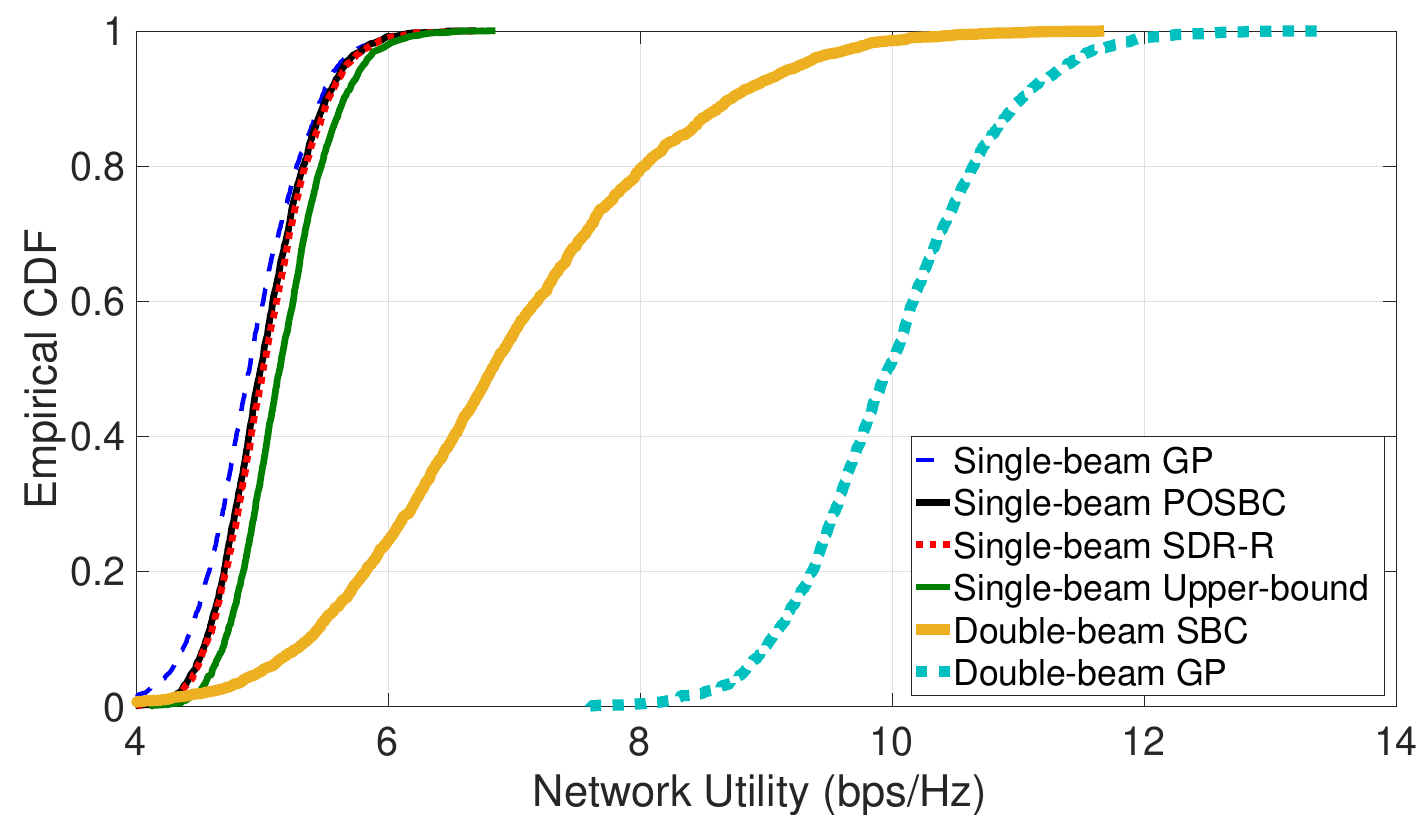}
 \caption{Empirical CDF of network utility}
 \label{fig:sim}
\end{figure}
    \begin{figure*}
    \centering
        \begin{subfigure}[b]{0.25\textwidth}
            \includegraphics[width=1\textwidth]            {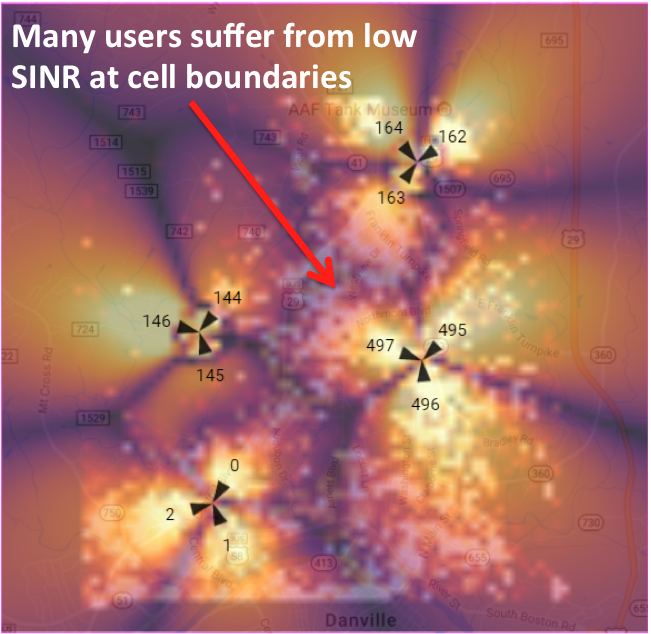}
            \caption[]%
            {{\small Network with passive antennas. Average throughput is 53.0 $\text{Mbps/km}^2$.}}    
            \label{fig:passive}
        \end{subfigure}
        \begin{subfigure}[b]{0.25\textwidth}              \includegraphics[width=\textwidth]{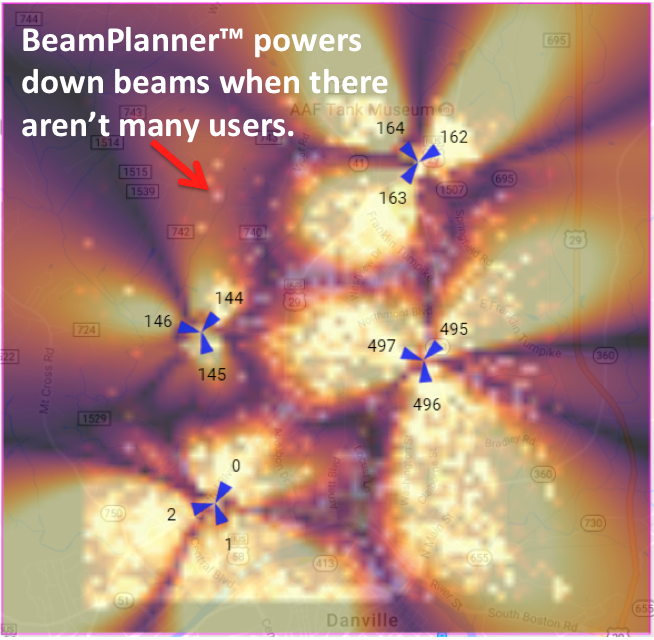}
            \caption[]%
            {{\small Network with optimized single beam phased arrays. Average throughput is 89.8 $\text{Mbps/km}^2$.}}    
            \label{fig:single_sector}
        \end{subfigure}
        \begin{subfigure}[b]{0.25\textwidth}   
            \centering 
            \includegraphics[width=\textwidth]{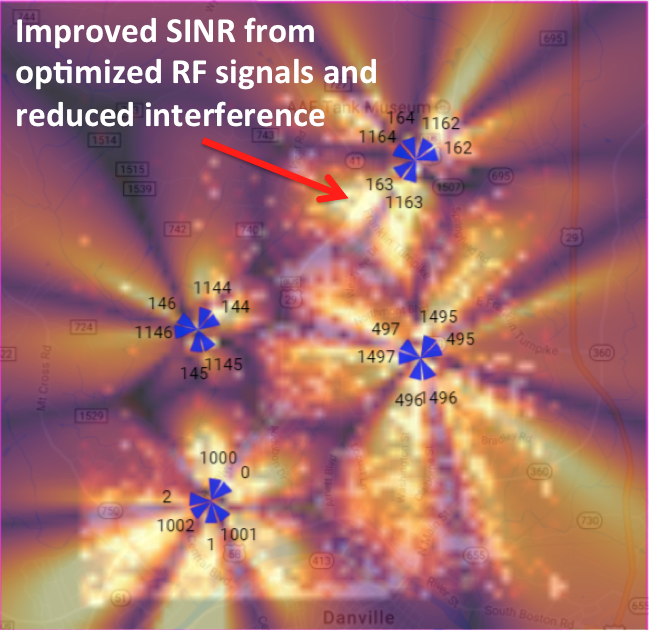}
            \caption[]%
            {{\small Network with optimized double beam phased arrays. Average throughput is 237.0 $\text{Mbps/km}^2$.}}
            \label{fig:dual_optimized}
        \end{subfigure}
        
        \caption{Sample cellular network located in Danville, VA optimized using $\text{BeamPlanner}^{\text{TM}}$ software. }
        \label{fig:beamplanner}
        \end{figure*}

\subsection{Performance evaluation}
In this section, we consider single-beam ($L=1$) and double-beam ($L=2$) scenarios. In order to compare the performance of the algorithms described in Sections \ref{sec:single-beam} and \ref{sec:multi-beam}, we consider 4000 random network realizations and calculate the network utility corresponding to each algorithm for each realization. For the single-beam scenario, the upper-bound of the network utility is obtained for each realization by solving problem $\Pi_{1r}$. It is assumed that $N_{trial}=10^3$ for SDR-R and SB-POSBC. Fig. \ref{fig:sim} illustrates the empirical CDF of network utility corresponding to each algorithm for $K=8$ hotspots. We observe that SDR-R outperforms SB-POSBC and GP in the single-beam scenario. Moreover, SB-POSBC performs very close to SDR-R. Having two beams will double the number of transmissions compared to the single-beam scenario which can potentially lead to significant network utility improvement if the interference due to multi-user activity (i.e. inter-beam interference) is managed appropriately. Since double-beam GP considers interference, it leads to almost 2X improvement in network utility compared to the single-beam algorithms. On the other hand, the performance of double-beam SBC is remarkably inferior to double-beam GP, due to lack of interference management.   
        
\section{Concluding Remarks}
\label{sec:conclusion}
We have studied the hybrid BF problem for a single macro-cell scenario where the BS is equipped with a massive phased array. Long-term channel information is used to design the BF vectors, where they are updated when there is a substantial change in long-term channel information. Several algorithms with different complexities have been proposed for designing BF vectors in different scenarios.

Based on the multi-cell generalization of the proposed algorithms, a commercial software has been developed by \emph{Blue Danube Systems} called \emph{$\text{BeamPlanner}^{\text{TM}}$}. The software is designed to optimize beam patterns in macro-cellular networks to enable effective antenna deployment. Fig. \ref{fig:beamplanner} illustrates the map of a sample cellular network in Danville, VA in three different deployment scenarios. Fig. \ref{fig:passive} represents the case where all cells are equipped with conventional passive antennas, whereas the other two figures showcase the deployment of \emph{$\text{BeamCraft}^{\text{TM}}$ 500}, an active antenna array designed and manufactured by \emph{Blue Danube Systems}. The white dots show the distribution of demand inside the network and the illuminated patterns illustrate the SINR at each point. Fig. \ref{fig:single_sector} shows the single-beam scenario where the beams are optimized using the GP algorithm. Fig. \ref{fig:dual_optimized} illustrates the same result for the double-beam scenario. It can be seen that double-beam active antenna arrays with optimal beam patterns can offer close to 5X throughput improvement over current systems with conventional antennas.   
\begin{appendices}

\section{Proof of Lemma \ref{lem:optimal-projection}}
\label{sec:app-lemma1}
Based on the definition of optimal projection in \eqref{eq:optimal-projection}, we have
\begin{align}
\bff{\hat{W}}=\argmin_{\bff{X}\in \mathcal{A}} ~d(\bff{X},\bff{W}), \label{eq:projection-optimization}
\end{align}
where, $d(\bff{X},\bff{W})\triangleq ||\bff{X}-\bff{W}||_{\text{F}}^2$ is convex in $\bff{X}$. Besides, set $\mathcal{A}$ is also convex, hence \eqref{eq:projection-optimization} is a convex optimization problem. Let us assume that for the coefficient representing antenna element $m$ and beam $s$, $X_{ms}=Z_{ms}e^{j\Phi_{ms}}$  with $X_{ms}\geq 0$ and $W_{ms}=R_{ms}e^{j\Psi_{ms}}$ with $R_{ms}\geq 0$. We can reformulate \eqref{eq:projection-optimization} as follows
\begin{align*}
&(\bff{\hat{Z}},\bff{\hat{\Phi}})=\argmin_{\{Z_{ms},\Phi_{ms}\}} \\
&\sum_{m\in[M]}\sum_{s\in [L]} \Big( R_{ms}^2+Z_{ms}^2-2R_{ms}Z_{ms}\cos(\Psi_{ms}-\Phi_{ms}) \Big),\\
&\text{subject to:}~~\forall m:\sum_{s\in [L]} Z^2_{ms}\leq1/M,~~ \forall m,s:~Z_{ms}\geq 0.
\end{align*}
Since $\forall m, s: Z_{ms}, R_{ms} \geq 0$, we have $\forall {m,s}: \hat{\Phi}_{ms}=\Psi_{ms}$. Then, the objective function is reduced to minimize $\sum_{m\in[M]} \sum_{s\in [L]}\big( R_{ms}-Z_{ms} \big)^2$. Furthermore, since all terms inside the sum are non-negative, the above problem can be broken into $M$ separate problems that can be solved independently for each antenna element index $m$. Hence, we can drop the antenna element index and rewrite the problem as follows:
\begin{align*}
&\min_{\{Z_s\}}\sum_{s\in[L]} \big( R_s-Z_s \big)^2  \\
&\text{subject to:} \quad \sum_{s\in[L]} Z^2_{s}\leq 1/M, \quad \forall s: Z_s \geq 0.
\end{align*}
It can be easily verified that the above problem is convex, therefore, the Karush-Kuhn-Tucker (KKT) conditions will result in the optimal solution \cite[chapter 5]{boyd2004convex}. First, we generate the Lagrangian as follows: 
\begin{align*}
&\mathcal{L}(\mathbf{Z}, \boldsymbol{\lambda}, \mu) = \\\nonumber &\sum_{s\in[L]} \big( R_s-Z_s \big)^2 + \mu\Big(\sum_{s\in[L]}Z_s^2 - 1/M\Big) - \sum_{s\in[L]}\lambda_sZ_s.
\end{align*}

KKT conditions to be checked are as follows:
\begin{align}
&\frac{\partial{\mathcal{L}}}{\partial{Z_s}} = 0~~~\forall s\in[L], \label{eq:c1}\\
& Z_s \geq 0 \qquad \forall s\in[L],\\
&\lambda_sZ_s=0~~~\forall s\in[L],\label{eq:c2}\\
&\mu, \lambda_s \geq 0 ~~~\forall s\in[L], \label{eq:c4}\\
&\sum_{s\in[L]} Z^2_{s}\leq 1/M,\\
&\mu\Big(\sum_{s\in[L]}Z_s^2 - 1/M\Big) = 0\label{eq:c3}.
\end{align}
From (\ref{eq:c1}), we can conclude:
\begin{equation}\label{eq:lambda}
\lambda_s = 2(Z_s - R_s) + 2\mu Z_s.
\end{equation}
By applying (\ref{eq:c3}), we know that either $\mu = 0$ or $\sum_{s\in{L}}Z_s^2 = 1/M$. If $\mu = 0$ (hence $\sum_{s\in[L]} Z^2_{s}\leq 1/M$), then from (\ref{eq:lambda}) we obtain $\lambda_s = 2(Z_s - R_s)$. According to (\ref{eq:c2}), we either have $\lambda_s = 0$ or $Z_s = 0$. If $Z_s = 0$, we will have $\lambda_s<0$ which contradicts (\ref{eq:c4}). Otherwise if $\lambda_s=0$, we have $Z_s = R_s$. 

On the other hand, if we assume $\mu>0$, then we have $\sum_{s\in[L]}Z_s^2 = 1/M$. Again, from (\ref{eq:c2}), if $\lambda_s > 0$, then $Z_s = 0$ and using \eqref{eq:lambda} we have $\lambda_s=-2R_s < 0$ which is a contradiction. If we set $\lambda_s=0$, we have $Z_s=R_s/(1+\mu)$ according to \eqref{eq:lambda}, which results in the following equalities:
\begin{align*}
&\sum_{s\in[L]}Z_s^2 = 1/M,\quad \sum_{s\in[L]}(R_s/(1+\mu))^2 = 1/M. 
\end{align*}
With a simple substitution we obtain $\mu = \sqrt{M\sum_{s=1}^LR_s^2}-1$ which yields $Z_s = R_s/\sqrt{M\sum_{s\in[L]}R_s^2}$. It should be noted that $\mu > 0$ and hence, $R_s > Z_s$. Therefore, this case represents the cases in which $\sum_{s\in[L]} R_s^2 \geq 1/M$.

Consequently, the solution of the problem is as follows:
\begin{align*}
  \hat{Z}_{ms}=\begin{cases}
    R_{ms} & \mbox{if } \sum_{s\in[L]} R_{ms}^2 < 1/M,\\
    R_{ms}/\sqrt{M\sum_{s\in[L]}R_{ms}^2} & \mbox{if } \sum_{s\in[L]} R_{ms}^2 \geq 1/M.
  \end{cases}
\end{align*}
for every $m\in[M]$ which concludes the proof. \sqproof  \notag



\end{appendices}

\bibliographystyle{IEEEtran}
\bibliography{reference}

\end{document}